\documentstyle[aps,12pt,epsfig,manuscript,amsbsy]{revtex}
\global\firstfigfalse
\begin{document}
\preprint{INJE-TP-00-1, hep-th/0001107}

\newcommand{\beq}{\begin{equation}}
\newcommand{\eeq}{\end{equation}}
\newcommand{\beqa}{\begin{eqnarray}}
\newcommand{\eeqa}{\end{eqnarray}}
\newcommand{\sr}{\sqrt}
\newcommand{\fr}{\frac}
\newcommand{\mn}{\mu \nu}
\newcommand{\G}{\Gamma}

\title{Randall-Sundrum Choice in the Brane World
}
\author{Y.S. Myung\footnote{E-mail address:
ysmyung@physics.inje.ac.kr}, Gungwon Kang\footnote{E-mail address:
kang@physics.inje.ac.kr}, and H.W. Lee\footnote{E-mail address:
hwlee@physics.inje.ac.kr}}
\address{
Department of Physics, Inje University,
Kimhae 621-749, Korea}
\maketitle
\begin{abstract}
We discuss the Randall-Sundrum (RS) choice for $h_{MN}$ in the brane-world. 
We begin with
the de Donder gauge (transverse-tracefree) including 
scalar($h_{55}$), vector($h_{5\mu}$) and tensor($h_{\mu\nu}$)
in five dimensions for comparison. 
One finds that $h_{55}=0$ and $h_{5\mu}=0$.
This leads to the RS choice.
It appears that the RS choice is so restrictive for the five 
massless states, whereas it is unique for describing the massive states. 
Furthermore, one can establish the stability of the RS solution with the RS
choice only. 
\end{abstract}
 
\newpage
\section{Introduction}
\label{sec-introduction}
Recently there has been much interest in the Randall-Sundrum 
brane-world\cite{Ran9905221}. 
The key idea of this model is that our universe may be a brane embedded in
higher dimensional space. A concrete model is a single 3-brane embedded 
in five-dimensional anti-de Sitter space (${\rm AdS}_5$). Randall and 
Sundrum have shown that the longitudinal part ($h_{\mu\nu}$) of the
metric fluctuations satisfy the Schr\"odinger-like equation with an 
attractive delta-function. As a result, the massless zero mode which 
describes the localized gravity on the brane was found. Furthermore, 
the massive modes lead to the correction to the Newtonian potential as 
of $V(r) = G_N \fr{m_1m_2}{r}(1+\fr{1}{r^2k^2})$. 

However, we point out that this has been done with the RS choice 
(a four-dimensional transverse-tracefree gauge). 
It seems that this choice is so restrictive that the RS model can 
describe the tensor fluctuation only. 
Furthermore, in order to 
have the well-defined theory on the brane, one has to consider the 
transverse parts of $h_{5\mu}, h_{55}$. More recently, Ivanov and 
Volovich\cite{Iva9912242} 
found that the equation for $h_{55}$ takes the 
Schr\"odinger-like equation with a repulsive delta-function. 
But their linearized equation is not correct.

In the massless and massive cases, 
$h_{55}$ is a four-dimensional scalar and $h_{5\mu}$ is
a four-dimensional vector. Hence it is not natural to set these
fields to be zero, as is shown in the RS choice. At the first sight,
the RS choice does not seem to be consistent with the massive states.
This is because $h_{55}$ and $h_{5\mu}$ belong to the physical fields
and these cannot be gauged away because the general covariance is broken. 
In both cases we choose the other
gauge such as the de Donder gauge (a five-dimensional 
transverse-tracefree gauge) instead of the RS choice in the beginning. 

In this paper, we find the correct linearized equation including 
$h_{5\mu}, h_{55}$. We point out the validity of the RS choice 
in describing the massless states as well as massive ones 
in the RS brane model. Also we discuss its connection to the 
stability of the RS solution.

\section{Perturbation Analysis}
\label{sec-perturbation}
We start with the Einstein equation with the bulk cosmological
constant $\Lambda$ and the brane tension $\tilde\sigma$
\beq
R_{MN}-\fr{1}{2}g_{MN}R = \Lambda g_{MN} 
    +\sigma \sqrt{g_{55}} g_{\mu\nu}
\delta^{\mu}_M \delta^{\nu}_N,  
\label{EEQ}
\eeq
which is derived from the action
\beq
I = \fr{1}{2} \int d^5x \sqrt{g_5} (R+2\Lambda )
 +\tilde\sigma\int d^4x \sqrt{g_4}. 
\label{Action}
\eeq
The RS solution is given by
\beq
\bar{g}_{MN} = H^{-2}\eta_{MN}
\eeq  
with $H=k|z|+1$ and $\eta_{MN} = {\rm diag}(+----)$. Further
$\Lambda = -6k^2$ and $\sigma=\tilde\sigma\delta(z)$ 
with $\tilde\sigma = 6k$. 
Here the capital
indices $M,N, \cdots$ are split into $\mu, \nu, \cdots $ 
(four-dimensions: $x^{\mu}$) and $5(x^5=z)$.  

After the conformal transformation of 
$g_{MN} = \Omega^2 {\tilde g}_{MN}$ with $\Omega = H^{-1}$,
let us introduce the perturbation 
\beq
{\tilde g}_{MN} = \eta_{MN} +h_{MN}.
\label{Fluctuation}
\eeq
Its linearized equation for Eq.(\ref{EEQ}) takes the form
\beqa
\mbox{} && \Box h_{MN} +3\fr{\partial_K H}{H} \eta^{KL} 
\left (\partial_N h_{KM} +\partial_M h_{KN} -\partial_K h_{MN} \right )
\nonumber  \\
\mbox{} && -\left (\fr{2\Lambda +2\sigma}{H^2}\right ) h_{55}\eta_{MN}
-\fr{2\sigma}{H^2} \left \{ h_{MN}
-\left ( h_{\mu\nu} + {h_{55}\over 2} \eta_{\mu\nu} \right )
           \delta^{\mu}_M \delta^{\nu}_N\right \} =0.
\label{FEQ}
\eeqa
Ivanov and Volovich in the version 2 of ref.\cite{Iva9912242} 
have missed the second line of Eq.(\ref{FEQ}). 
They in the version 3 have missed all of $\sigma$-dependent terms
but have included the $\Lambda$-dependent term.
This appears because the terms without 
$\partial$ arising from the LHS of Eq.(\ref{EEQ}) cannot be
cancelled against 
those$\left ( H^{-2} \left [ \Lambda h_{MN}+\sigma \left ( 
h_{\mu\nu} + {h_{55}\over 2} \eta_{\mu\nu} \right ) 
\delta^\mu_M \delta^\nu_N \right ]\right ) $  
from the RHS of Eq.(\ref{EEQ}). 
This line vanishes if $h_{MN}$ reduces to $h_{\mu\nu}$ with 
$h_{55}=h_{5\mu}=0$.

Here we use the de Donder gauge
\beq
\partial^M h_{MN} =0, \qquad \qquad h^P_P=0. 
\label{Donder1}
\eeq
This means that 
\beq
h^{\mu}_{\mu} = h_{55}, \qquad \partial^{\mu} h_{\mu 5} = 
\partial_5 h_{55}, \qquad \partial^{\mu} h_{\mu\nu} = 
\partial_5 h_{5\nu}. 
\label{Donder2} 
\eeq
 From Eq.(\ref{FEQ}) we obtain three equations,
\begin{eqnarray}
&& \left ( \Box - {{12 k^2} \over H^2} - 3 f \partial_5 \right ) 
             h_{55} =0 ,
\label{eqh55} \\
&& \left ( \Box - {{12 k} \over H^2} \delta(z) \right ) h_{5\mu} 
   - 3 f \partial_\mu h_{55} =0 ,
\label{eqh5mu} \\
&& \left ( \Box + 3 f \partial_5 \right ) h_{\mu\nu} 
  - 3 f \left ( \partial_\mu h_{5\nu} +\partial_\nu h_{5\mu} \right ) 
  + {12 \over H^2} \left ( k^2 - {k\over 2} \delta(z)\right ) h_{55}\eta_{\mu\nu} =0 
\label{eqhmunu} 
\end{eqnarray}
with $f = \partial_5 H/ H$.
Taking the trace of (\ref{eqhmunu}) and comparing it with 
Eq.(\ref{eqh55}), one finds that $h_{55}$ should vanish.
In deriving $h_{55}=0$, we use the de Donder gauge in (\ref{Donder2}).
With $h_{55}=0$, Eq.(\ref{eqh5mu}) becomes a decoupled equation.
In analyzing the perturbations, if one finds a decouple one, 
then one should solve it first.
And then one has to check its consistency with the 
remaining equation (\ref{eqhmunu}).
In order to solve Eq.(\ref{eqh5mu}) first, we introduce the separation of 
variables as
\begin{equation}
h_{5\mu}(x,z) = C_\mu(z) \psi_5(x).
\label{separation}
\end{equation}
Then Eq.(\ref{eqh5mu}) takes the form:
\begin{eqnarray}
&& \left ( \Box_4 + m_5^2 \right ) \psi_5(x) =0,
\label{eqpsi5} \\
&& C_\mu''(z) + \left \{ {{12 k}\over H^2} \delta(z) 
               + m_5^2 \right \} C_\mu(z) =0
\label{eqcmu}
\end{eqnarray}
with the gauge condition of $C_\mu(z) \partial^\mu \psi(x)=0$.
Here the prime($'$) means the differentiation with respect to 
its argument.
Now let us solve Eq.(\ref{eqcmu}) first.
This is exactly the case of ref.\cite{Gas74QP}.
The solution must satisfy the equation 
$C_\mu''(z) +m_5^2 C_\mu(z)=0$ at everywhere, except $z=0$. 
And then we assume its plane wave solution as
\begin{equation}
C_\mu(z) = A_\mu e^{-i m_5 |z|}; ~~
C_\mu^{z>0}(z)=A_\mu e^{-im_5 z}, ~~
C_\mu^{z<0}(z) = {\tilde A}_\mu e^{i m_5 z}.
\label{solcmu}
\end{equation}
We note that this solution keeps the reflection symmetry of the RS 
solution as $C_\mu(z') = C_\mu(z)$, under $z' \to -z$. 
The coefficients in front are the same $A_\mu = {\tilde A}_\mu$ 
because of the continuity of the wave function.
The derivative of $C_\mu(z)$ is no longer continuous 
because of the presence of the delta-function.
That is, one has
\begin{equation}
{{\partial C_\mu} \over {\partial z} } \Bigg |_{z=0^+} 
-{{\partial C_\mu} \over {\partial z} } \Bigg |_{z=0^-} 
= -12 k A_\mu,
\label{bc}
\end{equation}
which leads to
\begin{equation}
i m_5 =  6k
\label{m5}
\end{equation}
This admits the tachyonic mass of $C_\mu(z)$ as 
$m_5^2 = - 36 k^2 < 0$.
In other words, the normalizable bound-state 
solution to Eq.(\ref{eqcmu}) is allowed if 
its energy($m_5^2$) is negative.
As a check, $C^t_\mu(z) = A_\mu e^{-6 k |z|}$ satisfies
\begin{equation}
{C^t_\mu}''(z) + 12 k \delta(z) C^t_\mu(z) 
+m_5^2(=-36 k^2) C^t_\mu(z) =0.
\label{eqcheck}
\end{equation} 

But it remains to check whether this solution is or not 
consistent with Eq.(\ref{eqhmunu}).
Acting $\partial^\mu$ on Eq.(\ref{eqhmunu}) and using 
Eqs.(\ref{Donder2}) and (\ref{eqh5mu}), 
one gets the condition\cite{Vol00inprep}
\begin{equation}
\left ( \delta(z) C_\mu \right )' 
- 3 {\rm sgn}(z) \delta(z) C_\mu =0.
\label{condition}
\end{equation}
We note that ${\rm sgn}(z)\delta(z)$ is not well defined at
$z=0$ and thus one requires
\begin{equation}
C_\mu (0) =0.
\label{boundary}
\end{equation}
An alternative solution which satisfies 
Eqs.(\ref{eqcmu}) and (\ref{boundary}) is the plane wave as
Eq.(\ref{solcmu}) but $C^p_\mu(0)=0$,
\begin{equation}
C^p_\mu (z) = A_\mu \sin m_5 |z|.
\label{plane}
\end{equation}
At this stage, we remind the reader that our background is AdS$_5$ with 
$\delta(z)$-source.
This means that the solution to the linearized equations should 
carry at least the parameter `` $k$'' because the size of AdS$_5$ box 
is $1/k$ approximately and the brane tension is $\tilde\sigma =6k$.
However this plane wave solution misses ``$k$''.
This seems to be a solution for 5D Minkowski but not for AdS$_5$ background.
This is so because, due to the condition (\ref{boundary}) 
this does not account for the presence of the brane at 
$z=0$($12k\delta(z) C_\mu$-term in Eq.(\ref{eqcmu})) appropriately.
On the other hand, if $C_\mu(0)\ne 0$, $\delta(z) C_\mu(z)$ can be taken into 
account(as in our tachyonic solution $C^t_\mu$).
That is, there is no solution which satisfies both 
Eqs.(\ref{eqh5mu}) and (\ref{eqhmunu}).
Hence we are in a dilemma if $h_{5\mu}$ is truely a massive vector in 
the RS brane world.

Consequently, the tachyonic solution $C^t_\mu(z)$ is not a physical 
one because it is incompatible with the tensor equation (\ref{eqhmunu}).
As it stands, the presence of this solution says that $h_{5\mu}$ should 
be rejected to have a well-defined theory.
Fortunately the consistency with Eq.(\ref{eqhmunu}) leads to 
$h_{5\mu}=0$ on the whole space $z$ as in the RS choice.
Furthermore, the analysis for the massless case ($m_5=0$) 
in Eq.(\ref{eqcmu}) leads to $A_\mu=0$.
This implies that there is no massless vector state on the brane.
Hence it is obvious that $h_{5\mu}$ should not be a propagating vector 
in the RS background.
From now on we set $h_{5\mu}=0$.

\section{Massless States}
\label{sec-massless}
These states correspond to $\partial_5 h_{MN}=0$.
Before we proceed, we are willing to count the number of independent 
components of the graviton $h_{MN}$. For $D=5$ dimensions, a symmetric
tensor field $h_{MN}$ has $5(5+1)/2=15$ independent components, some of
which can be eliminated by the gauge conditions (\ref{Donder1}). 
This is $-(5+1)$. Further, after choosing the guage (\ref{Donder1}), 
there exists a residual gauge degrees of freedom as\cite{Wei72GC,Myu0001003}  
\beq
h'_{MN} = h_{MN} -\partial_M \xi_N -\partial_N \xi_M. 
\label{Residual}
\eeq
Notice that $h'_{MN}$ satisfy the de Donder gauge (\ref{Donder1}) 
provided that
\beq
\partial_M \xi^M =0, \qquad\qquad \Box \xi^M =0. 
\eeq
Thus $(5-1)$ are eliminated by our freedom. Hence the number of 
massless degrees of freedom in $D=5$ is 
\beq
\fr{5\cdot 6}{2} -(5+1)-(5-1) = 5.
\label{NDF}
\eeq

In order to see how $5$ is composed, let us consider the conventional 
Kaluza-Klein (KK) model\cite{Sal81ICTP211,Cas91SS}. This corresponds to 
$\Box h_{MN}=0$. Its massless bound state 
($\partial_5h_{MN}=0$) in $D=5$ dimensions can be described by  
\beq
\Box h_{MN} = J_{MN}
\eeq
with the external source $J_{MN}$. The general covariance of 
massless case can be represented as a source conservation law of
$\partial^N J_{MN} =0$ with $J^M_M=0$. In this case we choose 
a Lorentz frame as 
\beq
\partial_1 =\partial_4, \qquad \qquad \partial_2 =\partial_3 
=\partial_5 =0.
\label{LF}
\eeq
In this frame, the effective interaction reduces to the 
positive-definite form
\beq
{\cal L}^{\rm KK}_{\rm massless} = \fr{1}{4} h_{MN}J^{MN} 
=\fr{1}{4} \sum^{2}_{\lambda = -2} J_{-\lambda}\fr{1}{\partial_1^2
-\partial_4^2} J_{\lambda}, 
\label{JJ}
\eeq
where $\lambda$ refers to ${\rm O} (2)$ helicity and
\beqa
J_{\pm 2} &=& \fr{1}{2} (J_{22}-J_{33}) \pm i J_{23},  
\label{J2} \\
J_{\pm 1} &=& J_{52} \pm i J_{53} , 
\label{J1}  \\
J_0 &=& \sqrt{\fr{3}{2}} J_{55} .
\label{J0}
\eeqa
The terms in (\ref{JJ}) with (\ref{J2}), (\ref{J1}), and (\ref{J0}) 
describe the exchanges of spin-2 graviton, spin-1 photon, and 
spin-0 scalar.  
Here we have 2 components (spin-2), 2 (spin-1), and 1 (spin-0) and 
summing up these leads to 5 in (\ref{NDF}). 
According to the stability analysis\cite{Myu86PLB75,Ran83NPB491}, 
it is stable if each pole in (\ref{JJ}) is positive-definite.
Hence the KK model is classically stable.

Now let us consider the same issue using the RS choice as follows:
\beq
h_{55}=0, \qquad h^{\mu}_{\mu} =0, \qquad h_{5\mu}=0, 
\qquad \partial^{\mu} h_{\mu\nu} =0. 
\eeq
These eliminate $-10$ in $15$. 
Furthere we point out that there exists a $D=4$ residual gauge as
\beq
h'_{\mu\nu} = h_{\mu\nu} -\partial_{\mu} \xi_{\nu} 
-\partial_{\nu} \xi_{\mu}
\eeq
with 
\beq
\partial_{\mu} \xi^{\mu} =0, \qquad\qquad \Box_4 \xi^{\mu} =0.
\eeq
This eliminates $-3$ degrees of freedom (DOFs). 
Hence we have left 2 DOFs ($= 15-10-3$) in the RS choice. 
This is appropriate for describing the graviton $h_{\mu\nu}$ only.
Under this gauge, one finds  from Eq.(\ref{eqhmunu})
\begin{equation}
\Box_4 h_{\mu\nu} = J_{\mu\nu}
\label{eqsource}
\end{equation}
with the source relations
\beq
J_{55}=0, \qquad J^{\mu}_{\mu}=0, \qquad J_{5\mu}=0, \qquad 
\partial^{\mu} J_{\mu\nu} =0. 
\eeq
Using these, Eq.(\ref{JJ}) reduces   
\beq
{\cal L}^{\rm RS}_{\rm massless} = \fr{1}{4} \Big[ J_{-2}
\fr{1}{\partial^2_1 -\partial^2_4} J_2 
+ J_2 \fr{1}{\partial^2_1 -\partial^2_4} J_{-2} \Big] 
\label{RSmassless2}
\eeq
with $J_{\pm 2} = J_{22} \pm i J_{23} $.
As a result, the RS choice can describe the massless spin-2 modes 
of $h_{\pm 2}=h_{22} \pm ih_{23}$ only
as it can do best. One cannot find the vector and scalar fields.
Three modes of $h_{\pm 1}=h_{52} \pm ih_{53}$
and $h_0=\sqrt{3/2} h_{55}$ are missed, in comparison with the 
conventional KK model. 
Furthermore, it is shown that the RS
background is stable because ${\cal L}^{\rm RS}_{\rm massless}$ 
in (\ref{RSmassless2}) is positive-definite.

\section{Massive states}
\label{sec-massive}
In this case we start with the de Donder gauge in (\ref{Donder1}).
But it turns out that the set of perturbation equations 
(\ref{eqh55})-(\ref{eqhmunu}) become 
\beq
(\Box +3f\partial_5) h_{\mu\nu} =0, \qquad h_{55} =h_{5\mu} =0, 
\eeq
which corresponds to the RS massive case. 
At this stage, it is convenient to introduce the new variables
$h_{\mu\nu} = H^{3/2}(z) \hat{h}_{\mu\nu} (x, z) =H^{3/2}(z) 
\psi_h(z) \hat{\hat{h}}_{\mu\nu}(x)$. 
$\hat{\hat{h}}_{\mu\nu}$ corresponds to the canonical form of 
$h_{\mu\nu}$\cite{Csa0001033}. Then one finds 
\beq
\Big[ \fr{\Box_4}{2} + [ -\fr{\partial^2_5}{2} +V(z) ] \Big] 
\hat{h}_{\mu\nu} =0, 
\eeq
with 
\beq
V(z) = \fr{15k^2}{8H^2} -\fr{3k}{2H} \delta (z).
\eeq
Considering the equation of $(\Box_4 +m^2_h)\hat{\hat{h}}_{\mu\nu} =
J_{\mu\nu}$ with $J_{55}=J_{5\mu}=0$, we find the source conservation 
law as follows:
\beq
\partial^{\mu} J_{\mu\nu} =0 \qquad \qquad 
J^{\mu}_{\mu}=0 .
\eeq
Further the mass $m^2_h$ is determined 
by the equation\cite{Ran9905221} 
\beq
\Big[ -\fr{1}{2}\partial^2_5 +V(z) \Big] \psi_h (z) 
= \fr{1}{2} m^2_h \psi_h(z). 
\eeq
It was shown that $V(r)$ guarentees $m^2_h \ge 0$\cite{Csa0001033}.
This implies that these are no normalizable negative energy 
graviton modes.
In this case we choose a massive Lorentz frame in which
\beq
\partial_1=\partial_5 \qquad {\rm and} \qquad \partial_i =0
\qquad {\rm for} \qquad i=2,3,4.
\eeq
It follows that, in the neighborhood of the pole, the effective 
interaction reduces to 
\beq
{\cal L}^{\rm RS}_{\rm massive} = \fr{1}{4} J_{ij} \fr{1}
{\partial^2_1 +m^2_h} J_{ij},
\eeq
where $J_{ij}$ is a symmetric traceless tensor in three 
dimensions\cite{Sal81ICTP211}. 

One finds a massive tensor with 5 DOFs because $J_{ij}$ has 
5 ($=3\cdot 4/2 -1$) components. 
It is interesting
to ask how we can interpret this DOFs. This is clear from the fact
that in the massive case the global symmetry of spacetime is 
spontaneously broken\cite{Cas91SS}. The gauge parameters 
$\xi_{\mu}(x,z)$ and $\xi_5(x,z)$ in (\ref{Residual}) are associated
to spontaneously broken generators. The $h_{\mu\nu}$ with 2 DOFs
acquire mass $m^2_h$ by eating 2 DOFs of $h_{5\mu}$-vector and 1 DOF 
of $h_{55}$ scalar. Thus one finds a pure spin-2 massive particle 
with 5 DOFs. Explicitly, these are $h_{23}, h_{24}, h_{34}$, and 
other two satisfying $h_{22}+h_{33}+h_{44}=0$\cite{Myu86PLB75}. 
All these have positive-definite norm states. 
Hence all of the massive states in the RS model are classically stable.

\section{Discussion}
\label{sec-discussion}
We study the validity of the RS choice in the RS model.
For this purpose we start with the de Donder gauge.
Using the RS choice for the massive case, one finds 5 DOFs 
in $h_{\mu\nu}$. These all turn out to be the physical massive modes. 
Hence there remains no residual gauge symmetry.
In the massless case, we have three gauge degrees of freedom upon 
choosing the RS one. This corresponds to a residual gauge degrees 
of freedom. Using these, we can always find the massless spin-2 with 
2 DOFs; for example, see Ref.\cite{Myu0001003}. Hence we always have 
a localized gravity in a 3-brane.  

For the stability of the RS solution of 
$ds^2_{\rm RS} = H^{-2}\eta_{MN}dx^M dx^N$, we find that 
${\cal L}^{\rm RS}_{\rm massless}$ has positive norm states for
the RS choice, and also ${\cal L}^{\rm RS}_{\rm massive}$ is
positive-definite with the RS choice. This means that the RS choice
is regarded as a unique one to establish the stability of the RS solution. 

Finally, we comment on the Schr\"odinger-like equation 
(\ref{eqh5mu}) with $h_{55}=0$.
It may imply that the RS model is not classically stable because 
it has a tachyonic mass.
However, considering its consistency with Eq.(\ref{eqhmunu}), 
this puzzle can be resolved.
If the RS model makes sense, $h_{5\mu}=0$.

\section*{Acknowledgments}

We would like to thank Volovich for sending his paper to us.
This work was supported by the Brain Korea 21 Program, Ministry of 
Education, Project No. D-0025.

\end{document}